\documentclass{PoS}

\usepackage{hepnames}
\usepackage{booktabs}
\usepackage{cleveref}
\usepackage{physics}

\newcommand{\Brhotaunu}{\PB \rightarrow \Prho \Ptau \Pnut}
\newcommand{\Bomegataunu}{\PB \rightarrow \Pomega \Ptau \Pnut}
\newcommand{\RRho}{R_{\Prho}}
\newcommand{\ROmega}{R_{\Pomega}}
\newcommand{\redRRho}{\hat{R}_{\Prho}}
\newcommand{\redROmega}{\hat{R}_{\Pomega}}

\newcommand{\BtoRhoLNu}{\PB \rightarrow \Prho \Plepton \Pnu}
\newcommand{\BtoRhoTauNu}{\PB \rightarrow \Prho \Ptau \Pnu}
\newcommand{\BtoOmegaLNu}{\PB \rightarrow \Pomega \Plepton \Pnu}
\newcommand{\BtoOmegaTauNu}{\PB \rightarrow \Pomega \Ptau \Pnu}
\newcommand{\BtoVLNu}{\PB \rightarrow V \Plepton \Pnu}
\newcommand{\BtoVTauNu}{\PB \rightarrow V \Ptau \Pnu}

\newcommand\mat[1]{\mathbf{#1}}
\renewcommand\vec[1]{\mathbf{#1}}

\newcommand{\transpose}{\mathrm{T}}

\title{Precision predictions for $\Brhotaunu$ and $\Bomegataunu$ in the SM and beyond}

\ShortTitle{Precision predictions for $\PB \rightarrow \Prho \Ptau \Pnut$ and $\PB \rightarrow \Pomega \Ptau \Pnut$ in the SM and beyond}

\author{\speaker{Markus Tobias Prim}\\
	Karlsruhe Institute of Technology (KIT), Institute of Experimental Particle Physics (ETP),\\
    76131 Karlsruhe, Germany\\
    E-mail: \email{markus.prim@kit.edu}}
\author{Florian Urs Bernlochner\\
	Physikalisches Institut der Rheinischen Friedrich-Wilhelms-Universit\"at Bonn,\\ 53115 Bonn, Germany \\
	E-mail: \email{florian.bernlochner@uni-bonn.de}}
\author{Dean J. Robinson\\
	Ernest Orlando Lawrence Berkeley National Laboratory, University of California,\\ Berkeley, CA 94720, USA \\
	E-mail: \email{drobinson@lbl.gov}}

\abstract{We present new precision predictions for semitauonic decays involving $\Prho$ and $\Pomega$ final state mesons. These decay channels offer an interesting orthogonal probe to study the existing B anomalies in semitauonic transitions and are accessible with the Belle II experiment. The predictions are based on combining existing light-cone sum-rule calculations for the form factors with measured experimental spectra from the BaBar and Belle collaborations. This allows us to reliably extrapolate the light-lepton form factor predictions to large values of the four-momentum transfer squared, $q^2$, and in turn to derive precise predictions for $\RRho$ and $\ROmega$, the ratio of the total decay rates of $\Brhotaunu$ and $\Bomegataunu$ for $\Ptau$ final states with respect to light leptons in the SM. In addition, we investigate the impact of all four-fermi operators on the semitauonic $q^2$ spectra and these ratios.}

\FullConference{
European Physical Society Conference on High Energy Physics - EPS-HEP2019 -\\
			10-17 July, 2019\\
			Ghent, Belgium}

\begin{document}

\section{Introduction}
\begin{figure}
	\centering
	\includegraphics[scale=1]{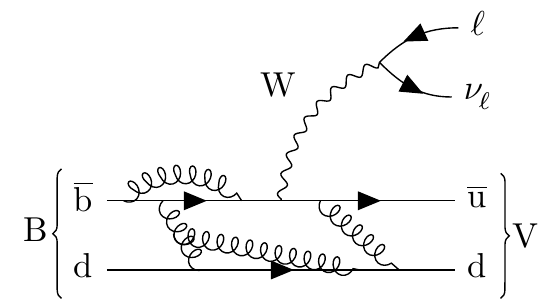}
	\quad \quad 
	\includegraphics[scale=1]{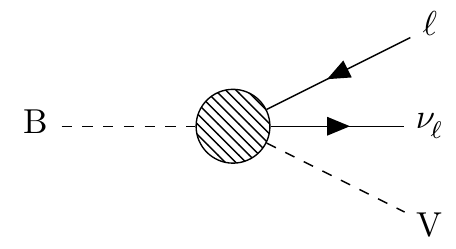}
	\caption{One possible parton level Feynman diagram (a) and the effective Feynman
		diagram (b). In the effective Feynman diagram, the propagator of the \PW is
		integrated out, i.e. the weak interaction is point-like, and the gluon interactions
		are described by the blob.}
	\label{fig:feynman-diagrams}
\end{figure}

The effective Standard Model (SM) Lagrangian describing semileptonic $\Pbottom \rightarrow \Pup \Plepton \Pnulepton$ transitions is given by
\begin{equation}
\mathcal{L_\mathrm{eff}} = \frac{-4G_\mathrm{F}}{\sqrt{2}} V_\mathrm{ub} (\APup \gamma_\mu P_\mathrm{L} \Pbottom) (\APneutrino \gamma^\mu P_\mathrm{L} \Plepton) + \mathrm{h.c.},
\end{equation}
with Fermi's constant $G_\mathrm{F}$, the CKM matrix element $V_\mathrm{ub}$ and the projection operator $P_\mathrm{L} = (1-\gamma_5)/2$. For resonant final states the hadronic matrix element for the $\Pbottom \rightarrow \Pup$ transition can be written as
\begin{equation}
\mel{V(p_V)}{ \APup \gamma^\mu P_\mathrm{L} \Pbottom }{\PB(p_{\PB})} = \sum T_i^\mu F_i(q^2),
\label{eq:tensor-decomposition}
\end{equation}
where $q=p_{\PB}-p_V$ is the four-momentum transfer in the decay and $V$ denotes any light final state meson (we will discuss $V \in \{\rho, \omega\}$). Further, the $T_i$ denote tensorial structures of the involved 4-momenta and polarizations, and the $F_i$ form factors. The sum runs over all allowed tensorial structures. 
The decay $\PB \rightarrow V \Plepton \Pnulepton$ is shown at parton level and as an effective diagram in Figure~\ref{fig:feynman-diagrams}.
In the right diagram the arms are described by the tensorial structures $T_i$ and the blob is described by the form factors $F_i$.
The form factors $F_i$ present in Equation~\ref{eq:tensor-decomposition} cannot be calculated with perturbation theory in the strong coupling constant and have to be determined using non-perturbative methods.

The Bourrely-Caprini-Lellouch (BCL) parametrization~\cite{Bourrely:2008za} is a model-independent ansatz for the form factors based on a fast converging series expansion of
\begin{equation*}
\begin{aligned}
z(q^2, t_0) &= \frac{\sqrt{t_+ - q^2} - \sqrt{t_+ - t_0}}{\sqrt{t_+ - q^2} + \sqrt{t_+ - t_0}} \, ,\\ 
\text{with}\quad t_+ &= (m_{\PB} + m_V)^2 \, ,\\
\text{and}\quad t_0 &= (m_{\PB} + m_V)(\sqrt{m_{\PB}} - \sqrt{m_V})^2 \, ,
\end{aligned}
\end{equation*}
where $m_{\PB}$ is the $\PB$ meson mass and $m_{V}$ is the mass of the final state meson. The form factors are expanded as:
\begin{equation*}
\begin{aligned}
F_i(q^2) &= P_i(q^2) \sum_k \alpha_k^i \left(z(q^2) - z(0)\right)^k \, ,\\
\text{with}\quad P_i(q^2) &= (1 - q^2/m_R^2)^{-1} \, ,
\end{aligned}
\end{equation*}
where $m_R$ is the mass of first resonance in the spectrum. For the transition into vector-like particles, i.e. $\PB \rightarrow V \Plepton \Pnulepton$, there exist a total of 8 independent form factors: $A_P$, $V$, $A_0$, $A_1$, $A_{12}$, $T_1$, $T_2$, $T_{23}$. The pseudoscalar form factor $A_P$ can be removed using the equations of motion:
\begin{equation*}
A_P = -2 \frac{m_M}{m_{\Pbottom} + m_{\Pup}} A_0 \, .
\end{equation*}
For the SM process, which is governed by a V-A current, only the vector ($V$) and axialvector ($A_0$, $A_1$, $A_{12}$) form factors contribute. Tensor currents can arise in models beyond the SM, and the three tensor ($T_1$, $T_2$, $T_{23}$) form factors can contribute.
It is convenient to express the V-A form factors in the helicity basis 
with the helicity amplitudes~\cite{NoComplexI}:
\begin{equation}
\begin{aligned}
H_\pm(q^2) &= \sqrt{\lambda(q^2)} \frac{V(q^2)}{m_{\PB} + m_V} \pm (m_{\PB} + m_V) A_1(q^2) \, ,\\
H_0(q^2) &= \frac{8m_{\PB}m_V}{\sqrt{q^2}}A_{12}(q^2) \, ,\\
H_s(q^2) &= \frac{\sqrt{\lambda(q^2)}}{\sqrt{q^2}}A_0(q^2) \, .
\end{aligned}
\end{equation}
The SM differential rate is then given by
\begin{equation}
\begin{aligned}
\frac{\mathrm{d}\Gamma}{\mathrm{dq^2}} &= \left| V_\mathrm{ub} \right| \frac{G_\mathrm{F}^2}{192 \pi^3 m_{\PB}^3} q^2 \sqrt{\lambda(q^2)} \left( 1- \frac{m_{\Plepton}^2}{q^2} \right)^2 \\
&\times \left[ \left( 1+\frac{m_l^2}{2q^2} \right)\left( H_+^2(q^2) + H_-^2(q^2) + H_0^2(q^2) \right) + \frac{3 m_l^2}{2q^2}  H_s^2(q^2) \right] \, ,
\end{aligned}
\label{eq:ulnu-theory:BToVlnuFull}
\end{equation}
with the Kaellen function $\lambda(q^2) = \left(\left(m_{\PB} + m_{V})^2 - q^2\right)\right)\left(\left(m_{\PB} - m_{V})^2 - q^2\right)\right) $, where $m_V$ denotes the mass of the final state meson. The differential rate can be simplified for light leptons $\Plepton = \Pe, \Pmu$ with the zero mass approximation $m_{\Plepton} = 0$:  
\begin{equation}
\frac{\mathrm{d}\Gamma}{\mathrm{dq^2}} =  V_\mathrm{ub}  \frac{G_\mathrm{F}^2}{192 \pi^3 m_{\PB}^3} q^2 \sqrt{\lambda(q^2)} \left[ H_+^2(q^2) + H_-^2(q^2) + H_0^2(q^2) \right] \, .
\label{eq:ulnu-theory:BToVlnu}
\end{equation}

The available theory calculations from light-cone sum-rules (LCSR) \cite{Straub:2015ica} for the coefficients of the BCL expansion are only valid in a $q^2$ region up to about $14\, \mathrm{GeV}^2$. For the $q^2 > 14\, \mathrm{GeV}^2$ region the prediction solely relies on extrapolation. In the following we show how the precision in the region $q^2 > 14\, \mathrm{GeV}^2$ can be improved by combining the theory prediction with experimental data.

\section{Data-Theory Fit}
\label{sec:data-theory-fit}
The values of the form factors $F_i$ are evaluated at three distinct points in the spectrum: $q^2 = [0, 7, 14]\, \mathrm{GeV}^2$ with the given BCL coefficients from the LCSR calculation. The resulting vector is named $\vec{f}_\text{LCSR}$. Their correlation $\mat{C}_\text{LCSR}$ is determined by generating random samples from a multivariate normal distribution using the covariance matrix from the LCSR calculation.
The associated $\chi^2_\text{LCSR}$ term is constructed as
\begin{equation}
\chi^2_\text{LCSR} = \Delta \vec{f}_\mathrm{t}^\transpose \mat{C}^{-1}_\text{LCSR} \Delta \vec{f}_\mathrm{t} \, ,
\end{equation}
with $\Delta \vec{f}_\mathrm{t} = \vec{f}(\vec{a}) - \vec{f}_\text{LCSR}$ denoting the vectorial difference involving a prediction vector $\vec{f}(\vec{a})$ depending on the expansion coefficients $\vec{a}$.
The $\chi^2$ terms for the individual experimental measurements of References~\cite{Sibidanov:2013rkk,delAmoSanchez:2010af,Lees:2012mq} are constructed as
\begin{equation}
\chi^2_\mathrm{exp}(\vec{a}) = \Delta \Gamma_\mathrm{e}^\transpose C_\mathrm{exp}^{-1} \Delta \Gamma_\mathrm{e} \, ,
\end{equation}
with $\Delta \Gamma_\mathrm{e} = \Gamma(\vec{a}) - \Gamma_\text{exp}$ denoting the vectorial difference of the predicted and measured rates.
The global $\chi^2$ function takes the form
\begin{equation}
\chi^2(\vec{a}) = \chi^2_\text{LCSR}(\vec{a}) + \sum_\mathrm{exp} \chi^2_\mathrm{exp}(\vec{a}) \, .
\end{equation}

\begin{figure}
\centering
\includegraphics[width=0.3\linewidth]{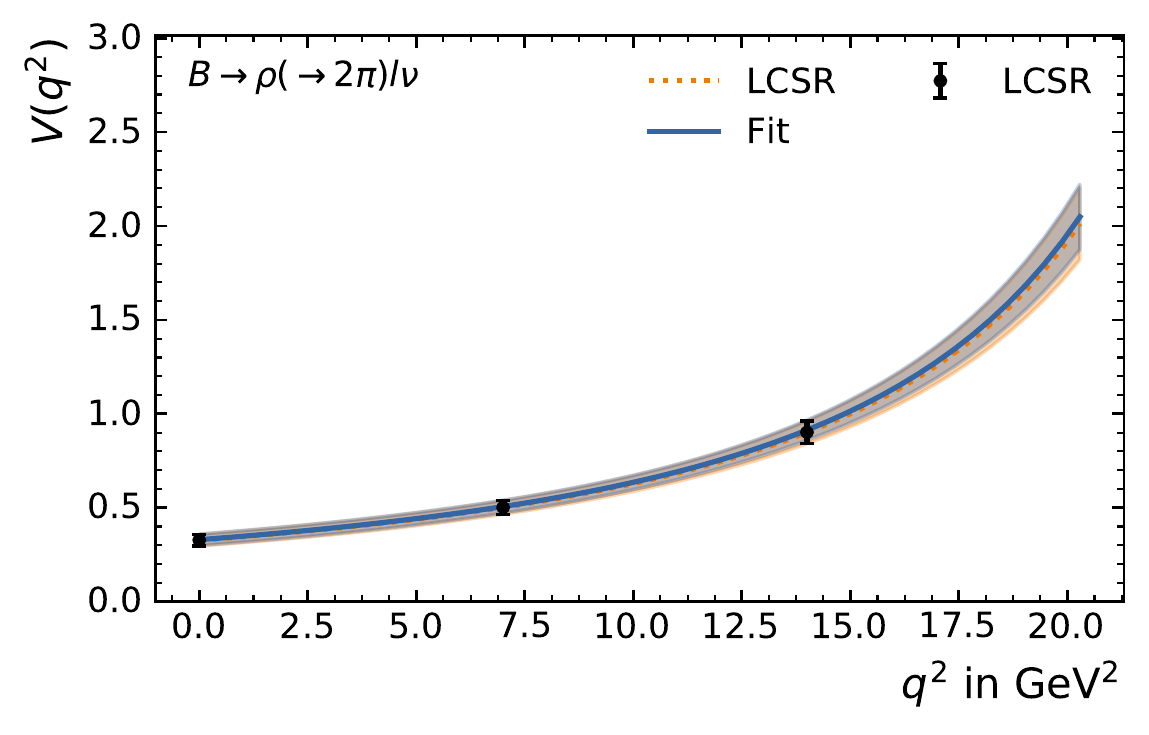}
\includegraphics[width=0.3\linewidth]{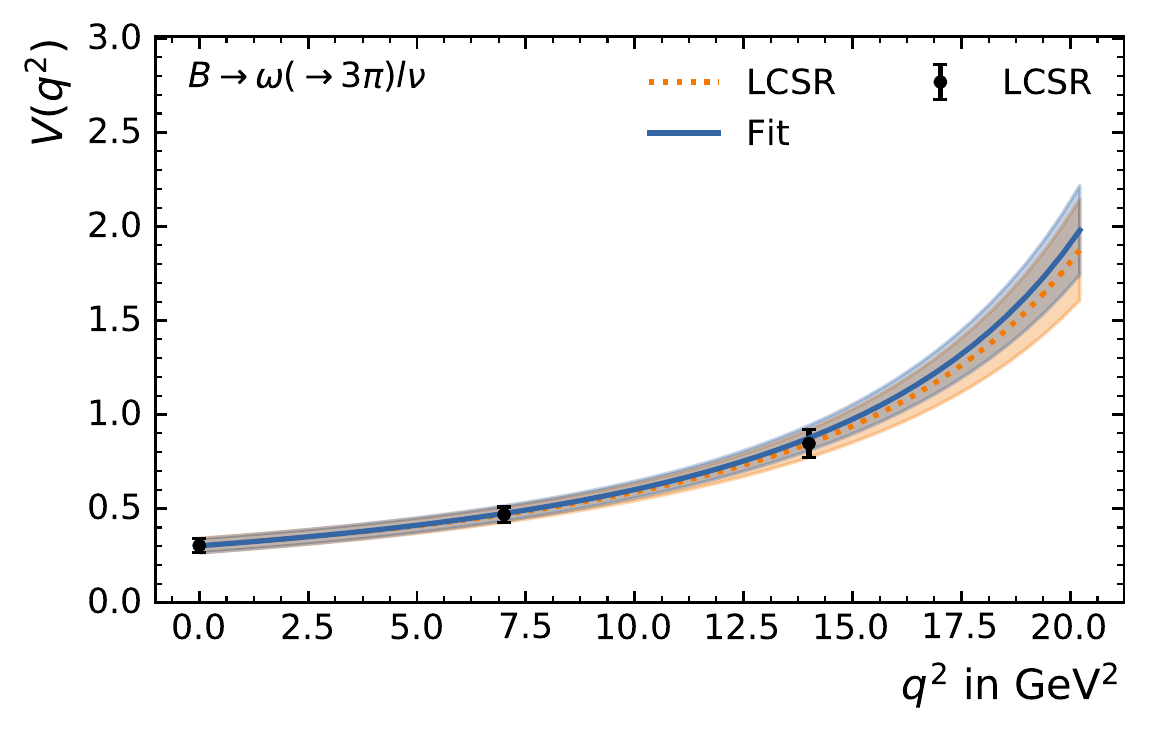}\\
\includegraphics[width=0.3\linewidth]{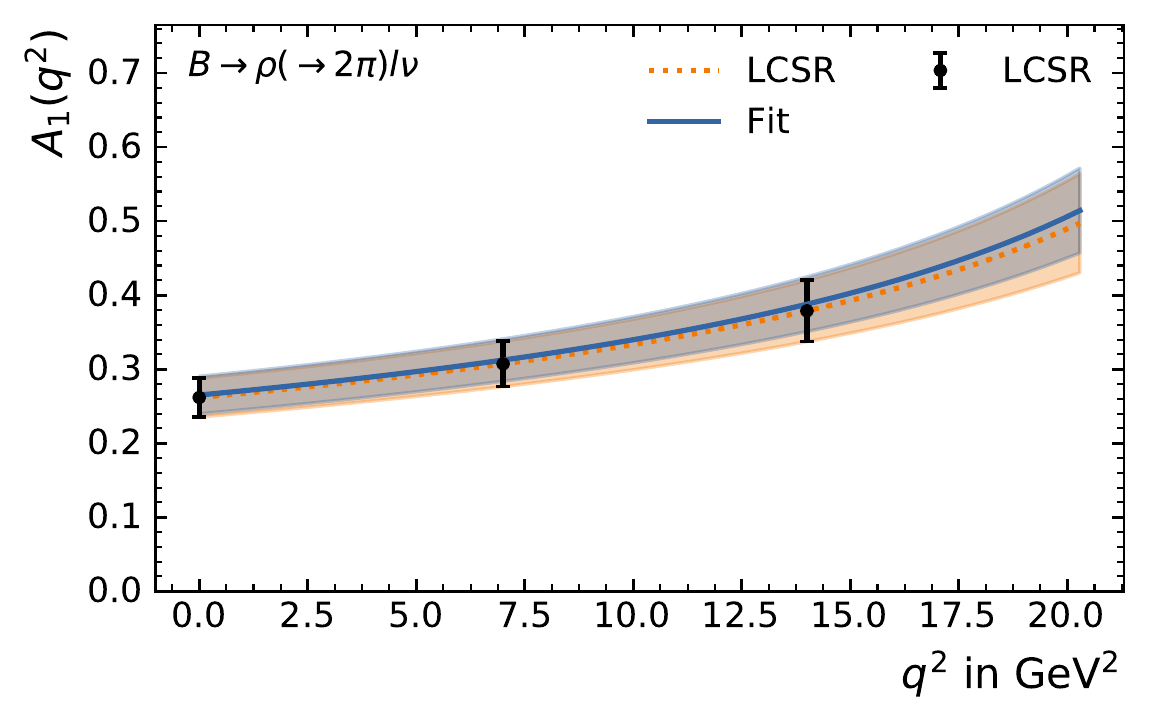}
\includegraphics[width=0.3\linewidth]{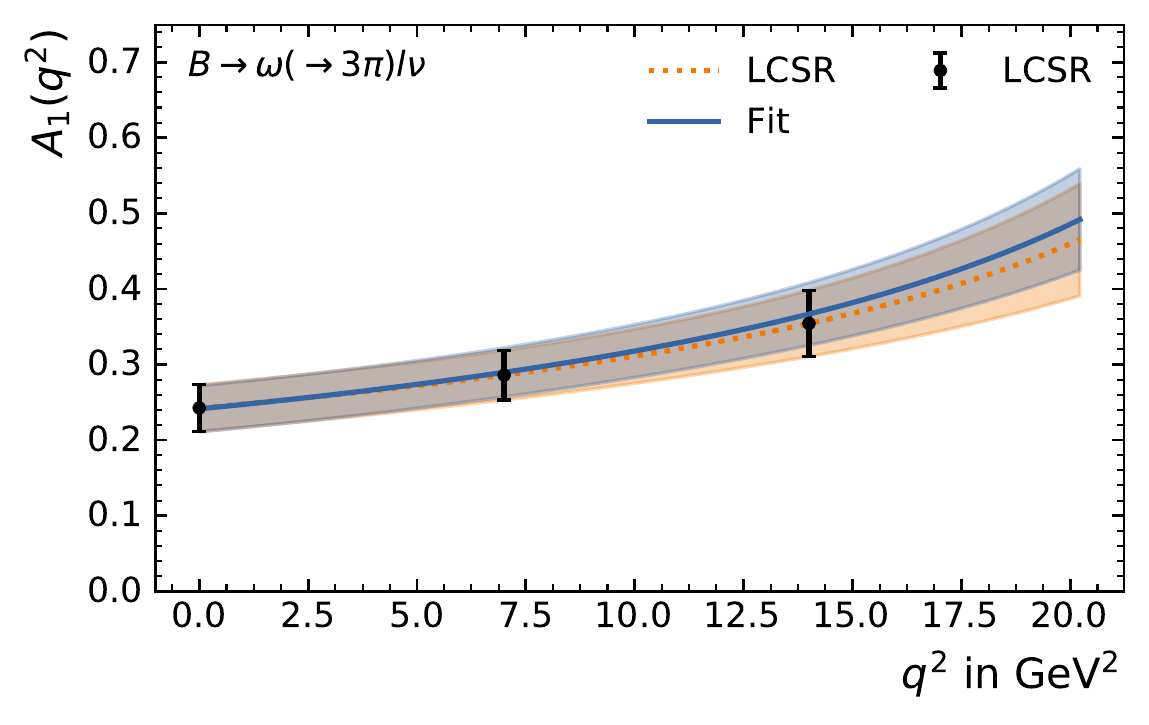}\\
\includegraphics[width=0.3\linewidth]{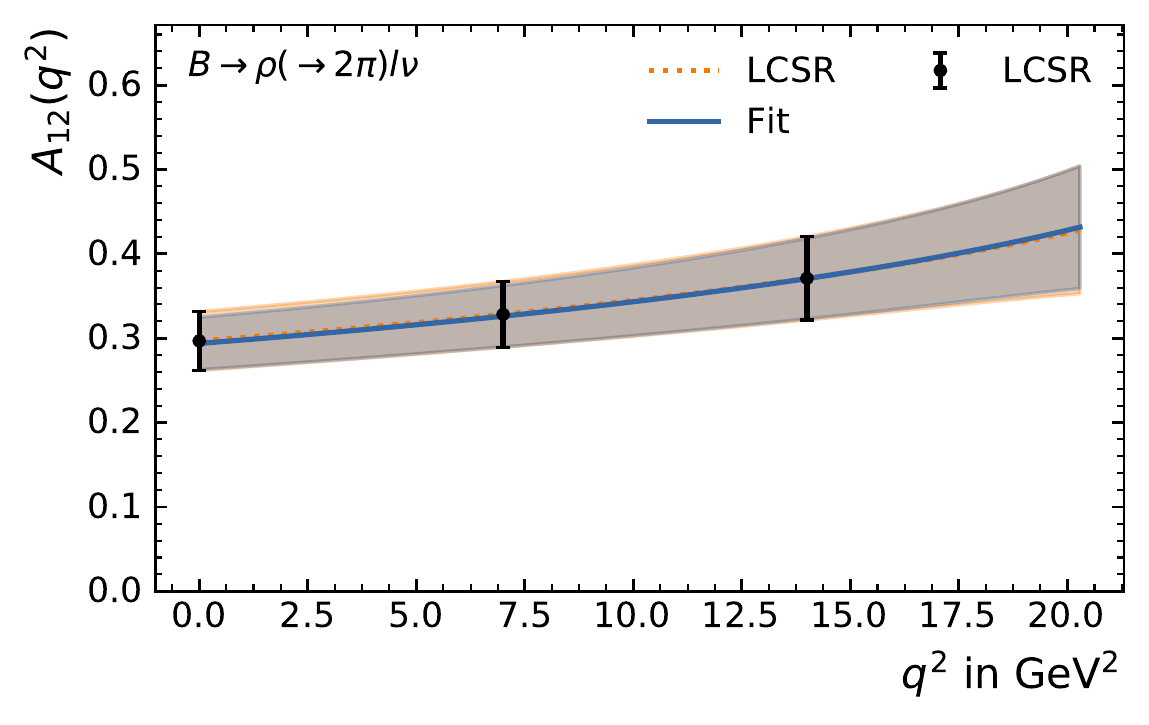}
\includegraphics[width=0.3\linewidth]{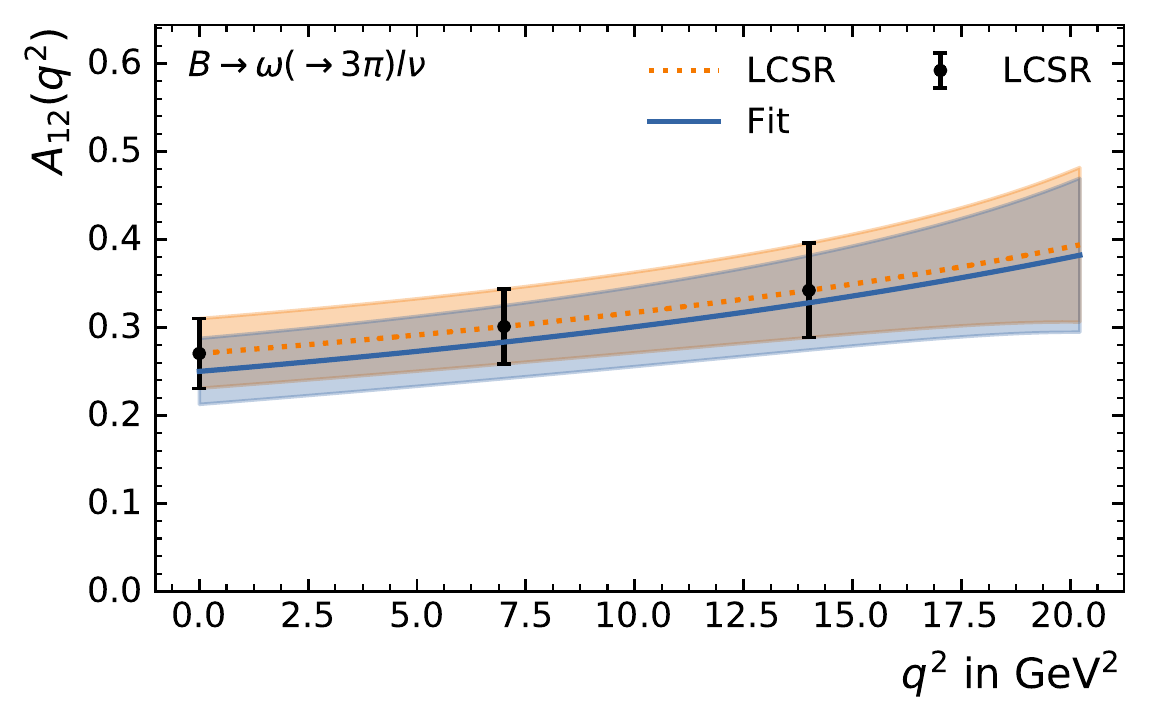}\\
\includegraphics[width=0.3\linewidth]{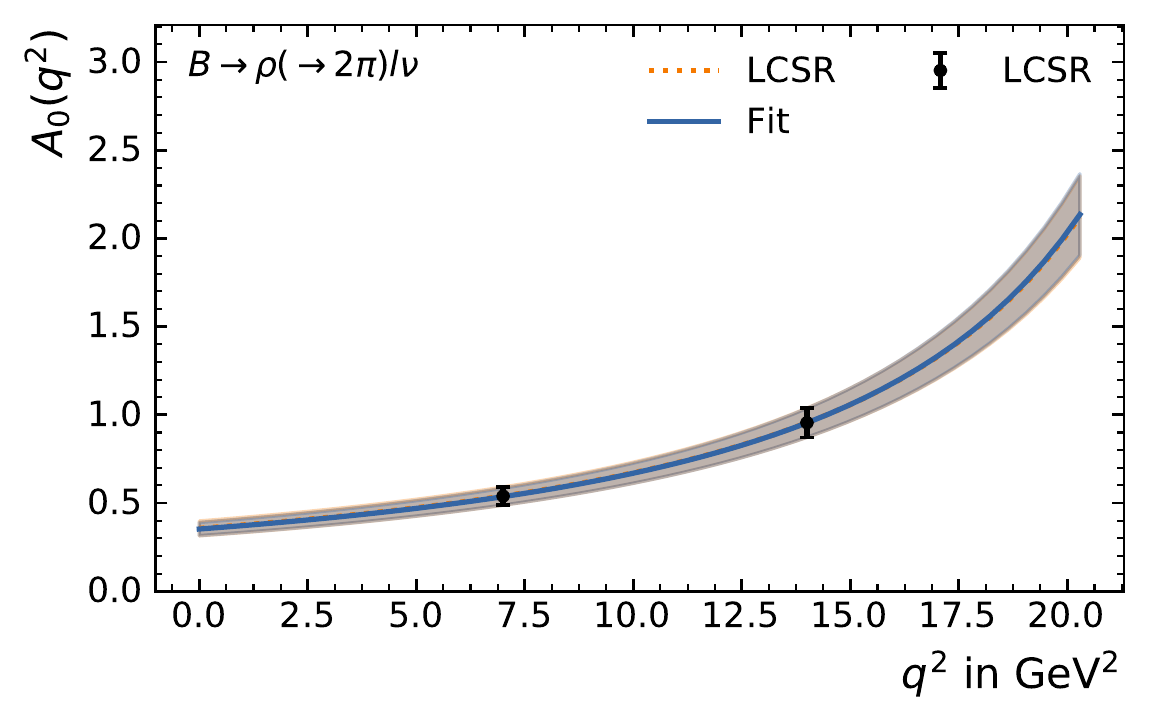}
\includegraphics[width=0.3\linewidth]{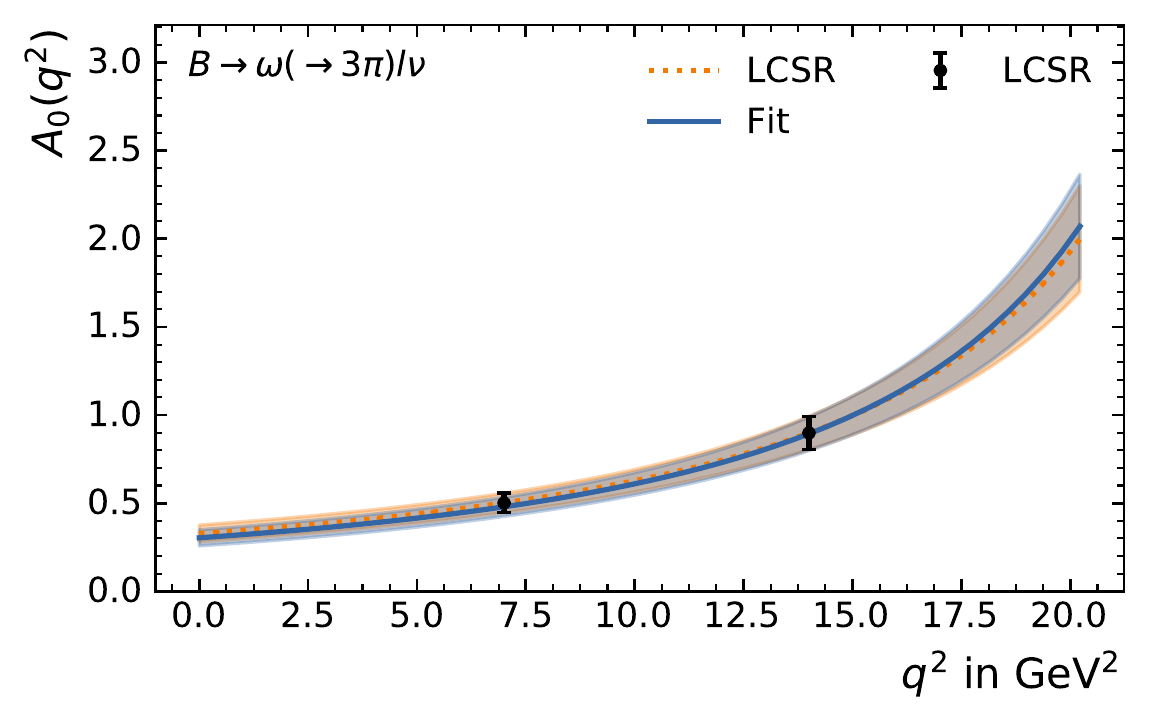}\\
\includegraphics[width=0.3\linewidth]{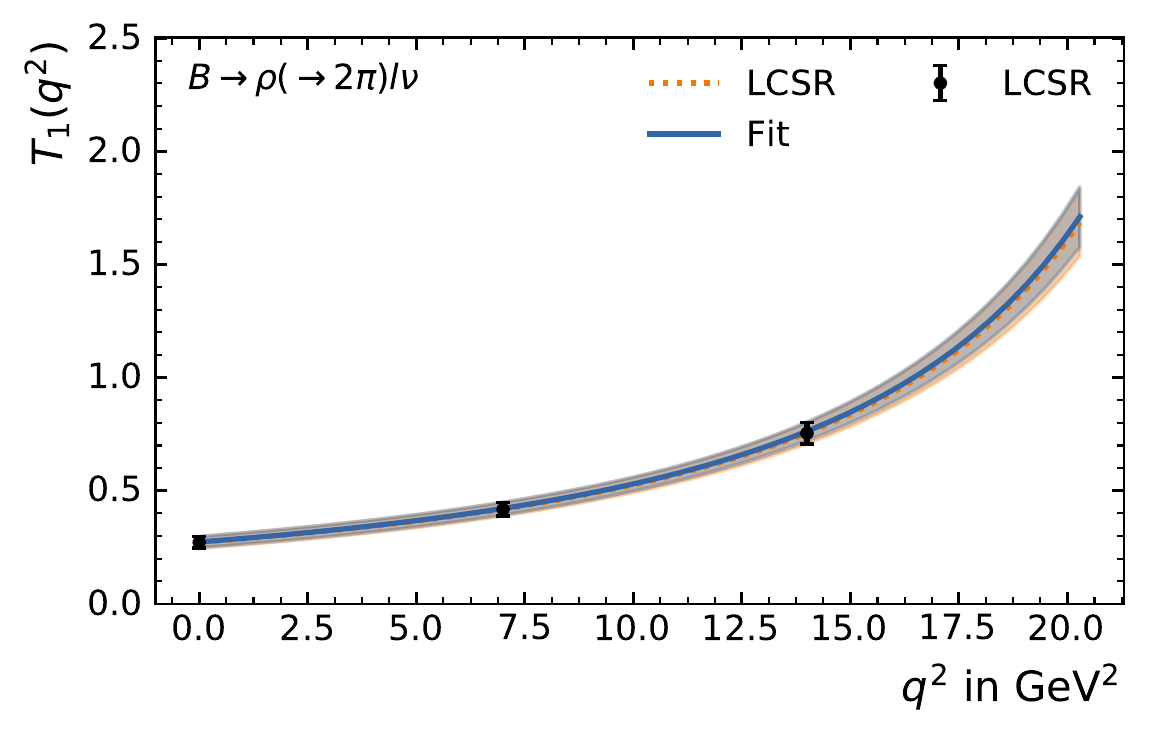}
\includegraphics[width=0.3\linewidth]{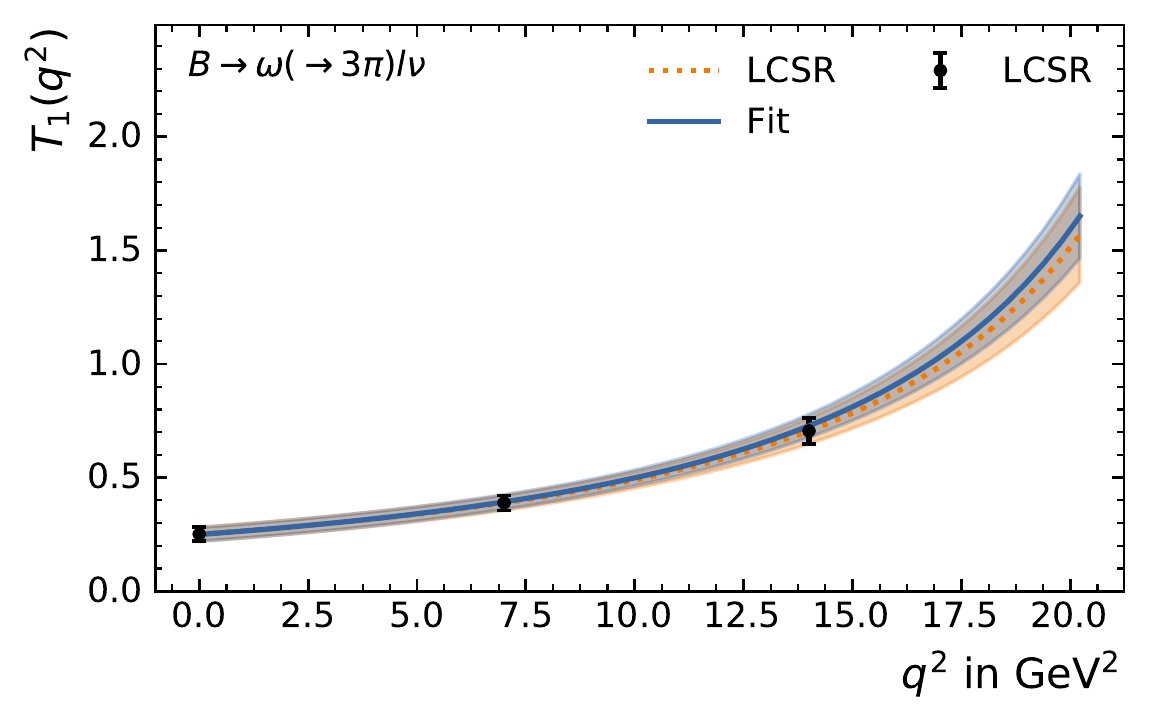}\\
\includegraphics[width=0.3\linewidth]{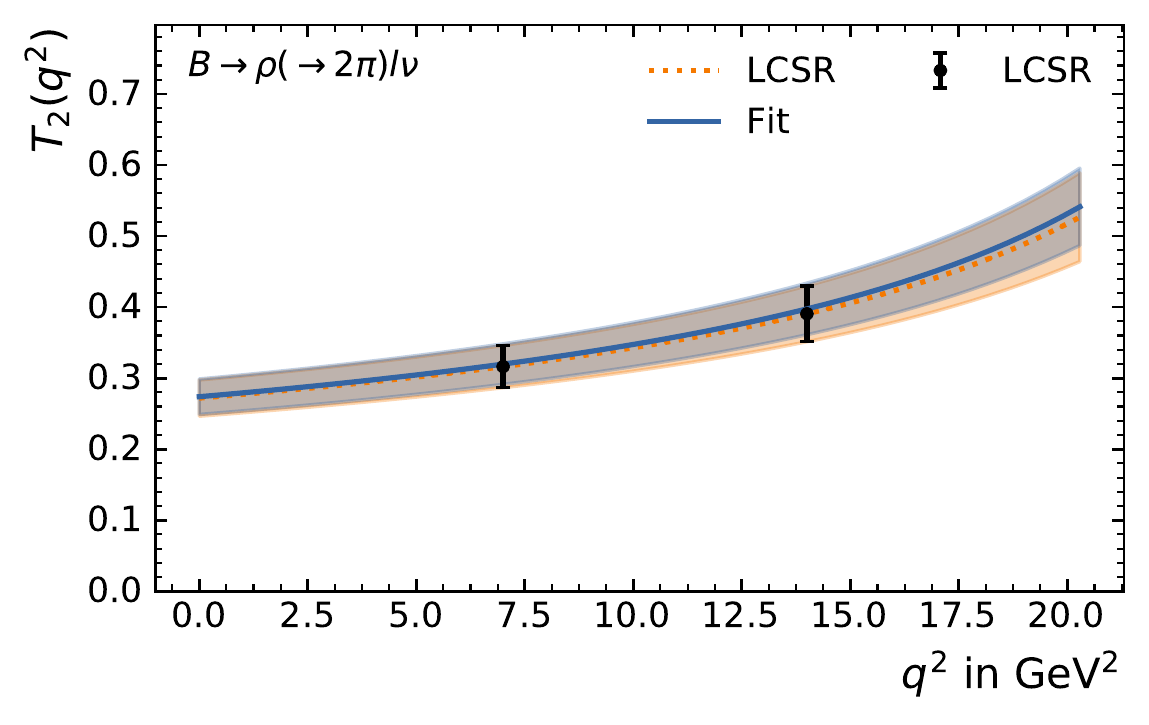}
\includegraphics[width=0.3\linewidth]{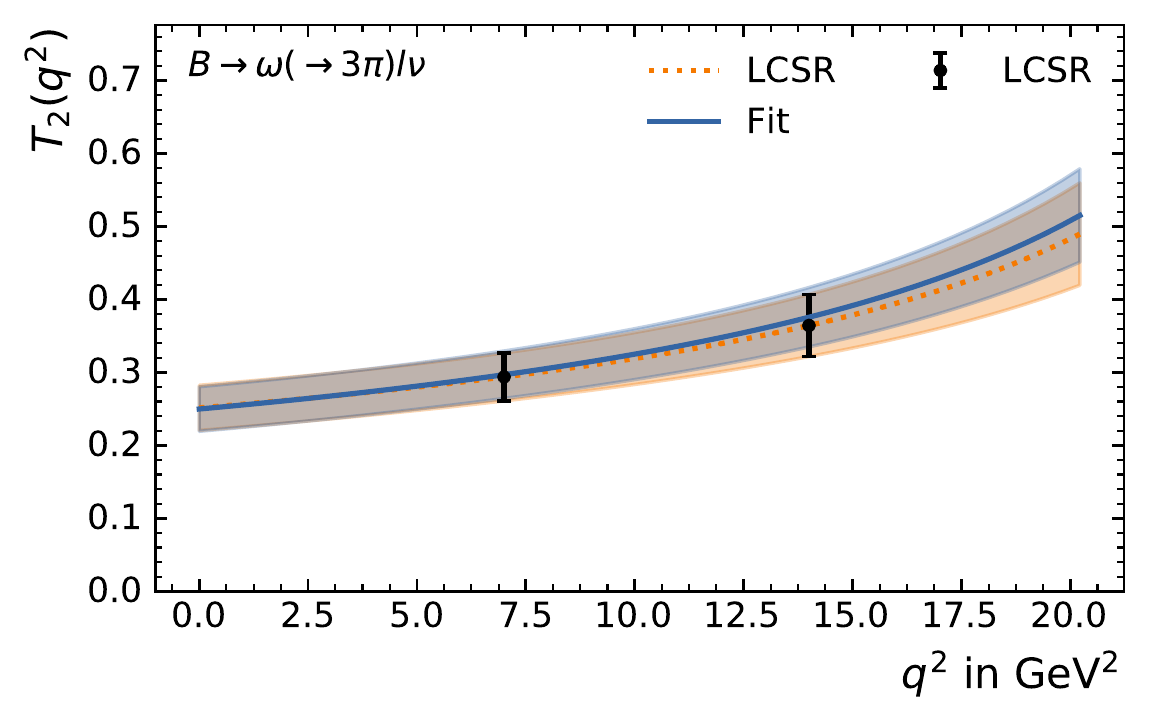}\\
\includegraphics[width=0.3\linewidth]{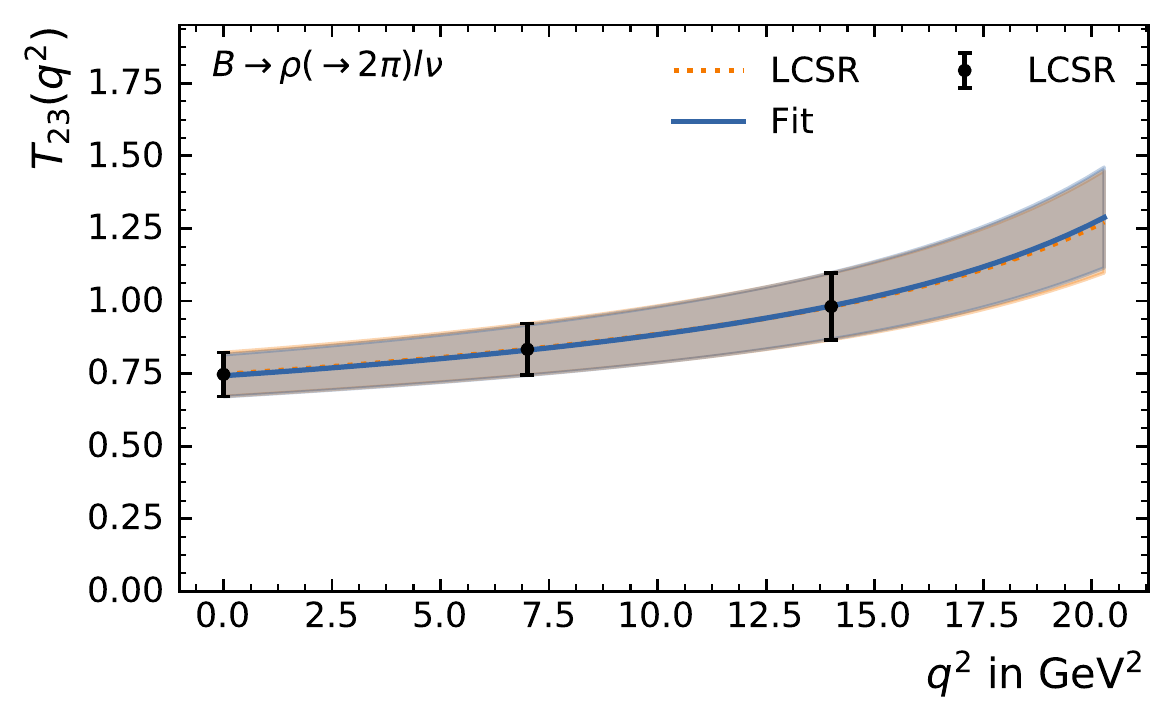}
\includegraphics[width=0.3\linewidth]{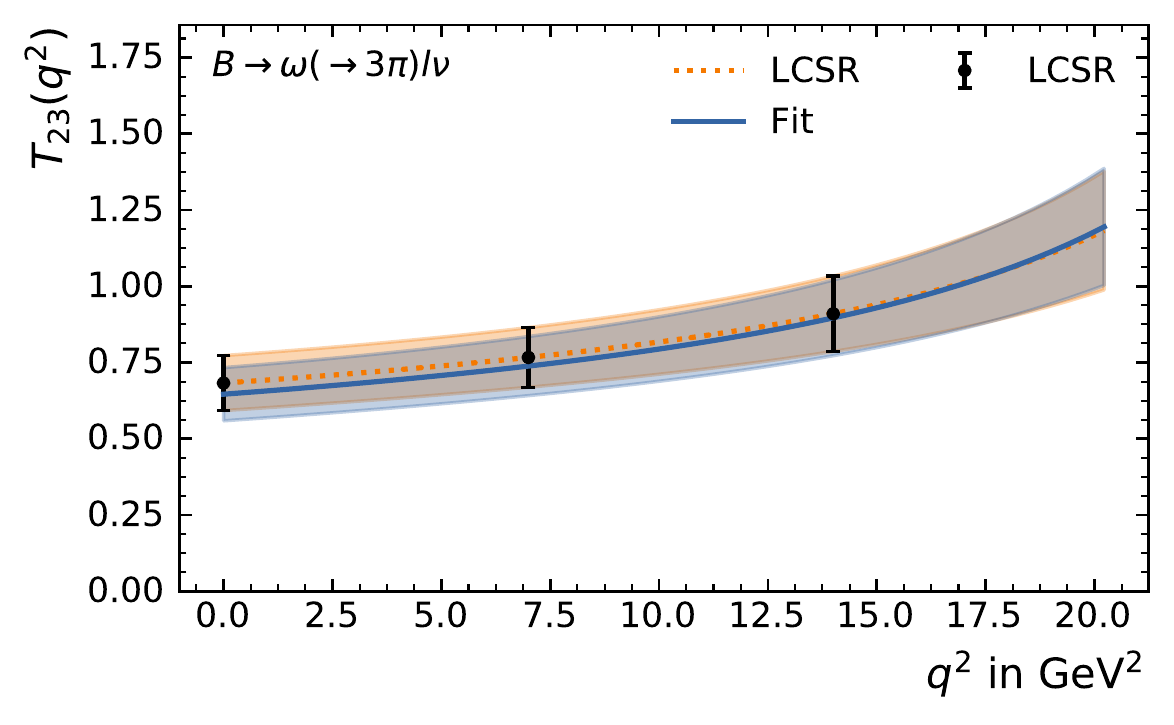}
\caption{Pre- and post-fit form factors. It is clearly visible that corrections on the form factors can be extracted from a combined theory and data fit.}
\label{fig:fit-results-ff}
\end{figure}

The form factors pre- and post-fit are given in Figure~\ref{fig:fit-results-ff}. It is clearly visible that corrections on the form factors can be extracted from the experimental measurements.

\begin{figure}
\centering
\includegraphics[width=0.45\textwidth]{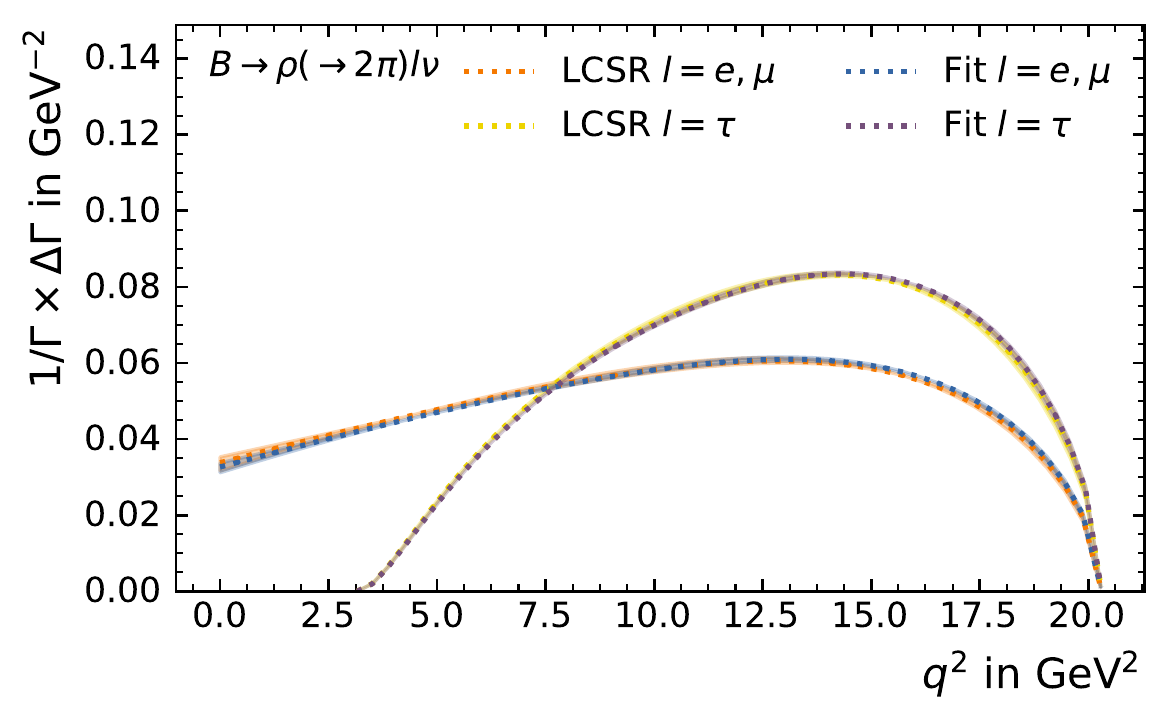}
\includegraphics[width=0.45\textwidth]{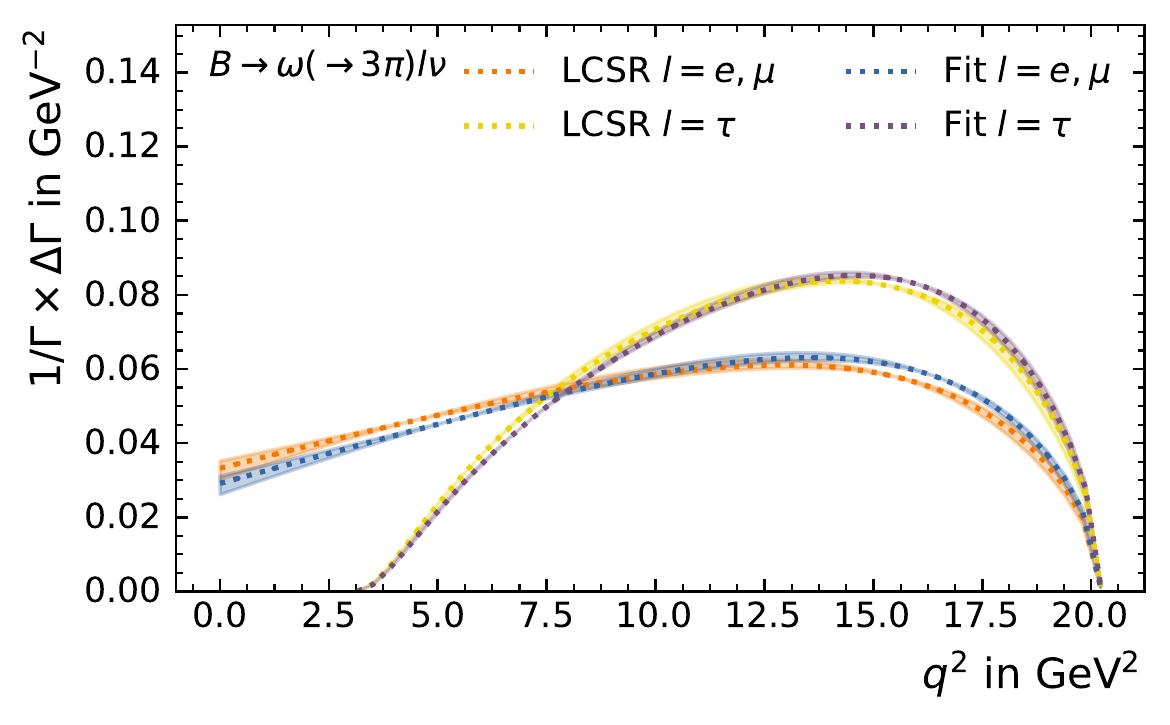}
\caption{The normalized $q^2$ spectra for $\BtoRhoLNu$ and $\BtoRhoTauNu$ (left) and $\BtoOmegaLNu$ and $\BtoOmegaTauNu$ (right) with the form factor expansion coefficients from light cone sum rules (dotted orange and yellow) and with the expansion coefficients extracted from the theory plus data fit (dotted blue and purple). }
\label{fig:fit-results}
\end{figure}

\section{Predicting $\RRho$ and $\ROmega$}

\subsection{SM Prediction}
We predict $\RRho$ and $\ROmega$ in two different $q^2$ intervals:
\begin{itemize}
	\item $q^2 \in [0, q^2_\mathrm{max}]\, \mathrm{GeV}$. The ratio is independent of the shape of the differential distributions.
	\item $q^2 \in [m_{\Ptau}^2, q^2_\mathrm{max}]\, \mathrm{GeV}$. The evaluation of the differential rate in the same $q^2$ region for tauonic and semileptonic final states increases the correlation between nominator and denominator in the ratio and thus larger cancellation of uncertainties is possible. However, the prediction then depends on the actual shape of the light-lepton differential rate, due to the introduced cut.
\end{itemize}
We define $R$ and $\hat{R}$ as:
\begin{equation*}
\begin{aligned}
	R_V &= \frac{\int_{m_\tau^2}^{q^2_\mathrm{max}} \frac{\mathrm{d}\Gamma(\BtoVTauNu)}{\mathrm{d}q^2}\, \mathrm{d}q^2}{\int_{0}^{q^2_\mathrm{max}} \frac{\mathrm{d}\Gamma(\BtoVLNu)}{\mathrm{d}q^2}\, \mathrm{d}q^2} \, , \\
	\hat{R}_V &= \frac{\int_{m_\tau^2}^{q^2_\mathrm{max}} \frac{\mathrm{d}\Gamma(\BtoVTauNu)}{\mathrm{d}q^2}\, \mathrm{d}q^2}{\int_{m_\tau^2}^{q^2_\mathrm{max}} \frac{\mathrm{d}\Gamma(\BtoVLNu)}{\mathrm{d}q^2}\, \mathrm{d}q^2} \, .
\end{aligned}	 
\end{equation*}

The predictions for $R_{V}$ and $\hat{R}_{V}$ are tabulated in Table~\ref{tab:sm-rrho-romega}.

\begin{table}
\centering
	\begin{tabular}{cccc}
	\toprule
	\midrule
	$R_V$ & LCSR & Fit & Improvement \\
	\midrule
	$\RRho$      & $0.532 \pm 0.011$ & $0.535 \pm 0.008$ & $25\,\%$           \\
	$\ROmega$    & $0.534 \pm 0.018$ & $0.546 \pm 0.015$ & $16\,\%$           \\
	\midrule
	$\redRRho$   & $0.605 \pm 0.007$ & $0.606 \pm 0.006$ & $\phantom{0}6\,\%$ \\
	$\redROmega$ & $0.606 \pm 0.012$ & $0.612 \pm 0.011$ & $\phantom{0}7\,\%$ \\
	\midrule  
	\bottomrule
	\end{tabular}
\caption{The predictions for the ratio of tauonic to leptonic final states. The column `LCSR' uses only the theory prediction for the form factors from~\cite{Straub:2015ica}. The column `Fit' uses the coefficients extracted from the fit described in Section~\ref{sec:data-theory-fit}. }
\label{tab:sm-rrho-romega}
\end{table}

\subsection{New Physics Contribution to $\RRho$ and $\ROmega$}
\begin{figure}
\centering
\includegraphics[width=0.45\textwidth]{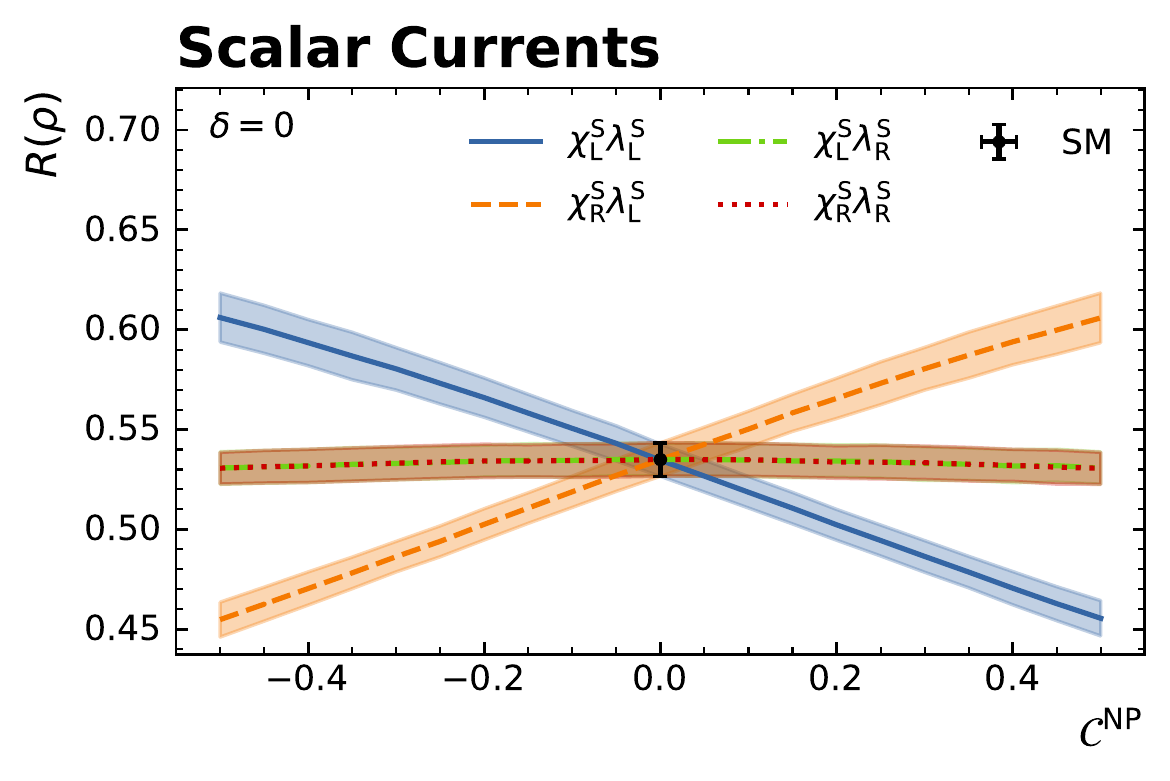}
\includegraphics[width=0.45\textwidth]{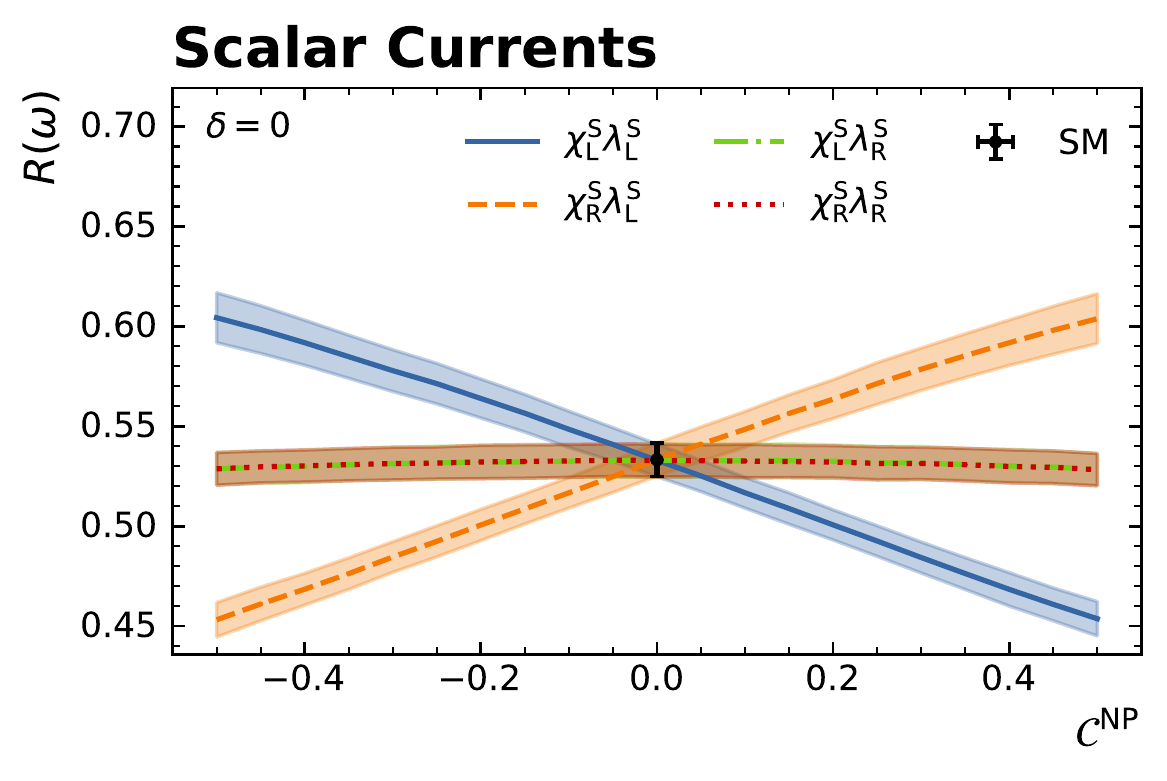}\\
\includegraphics[width=0.45\textwidth]{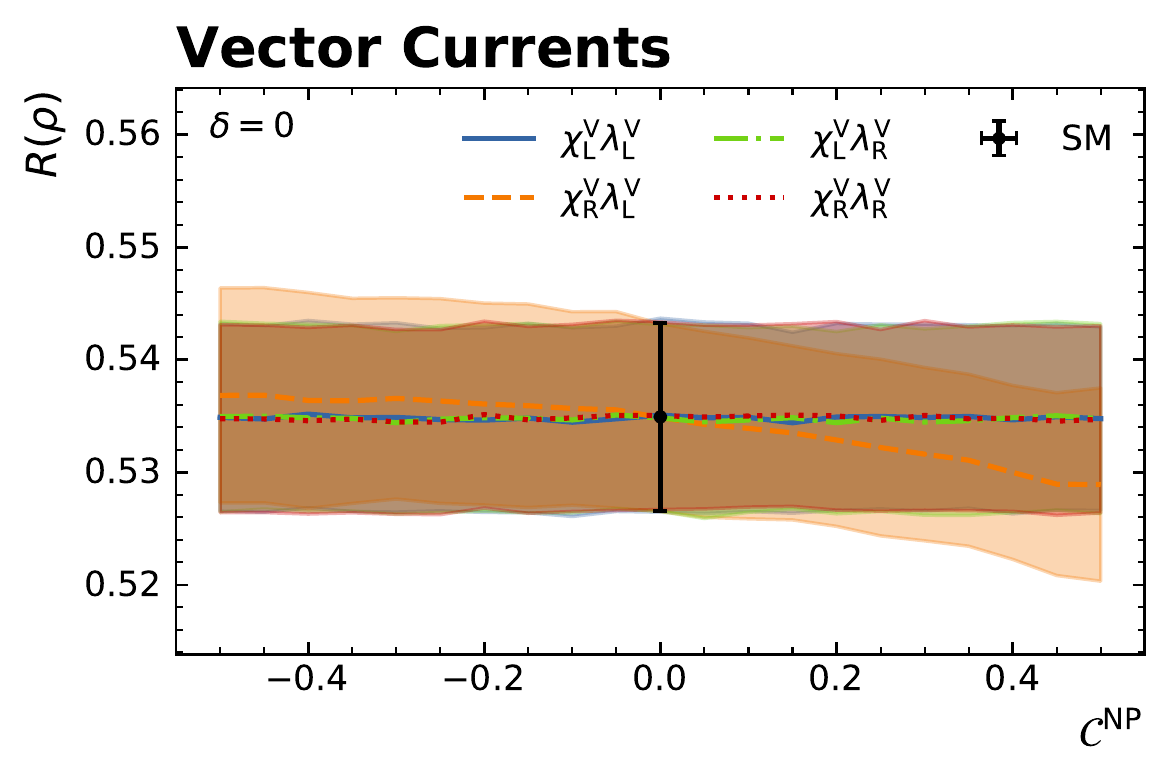}
\includegraphics[width=0.45\textwidth]{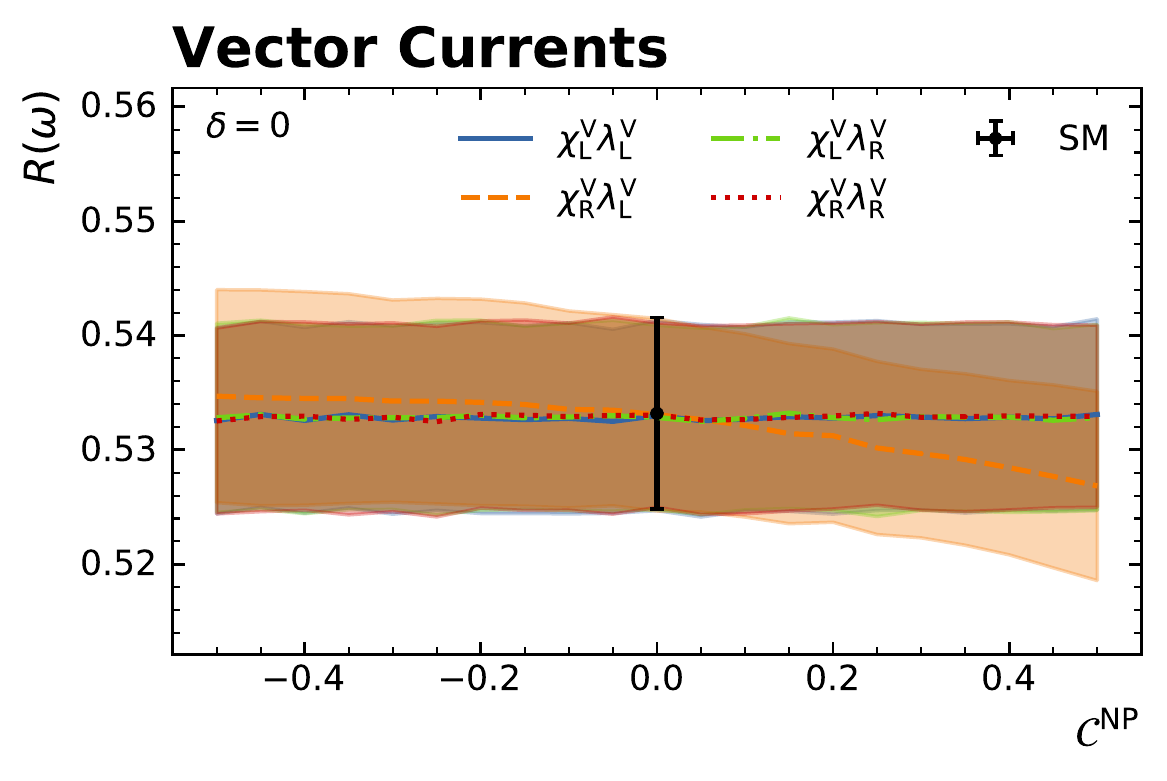}\\
\includegraphics[width=0.45\textwidth]{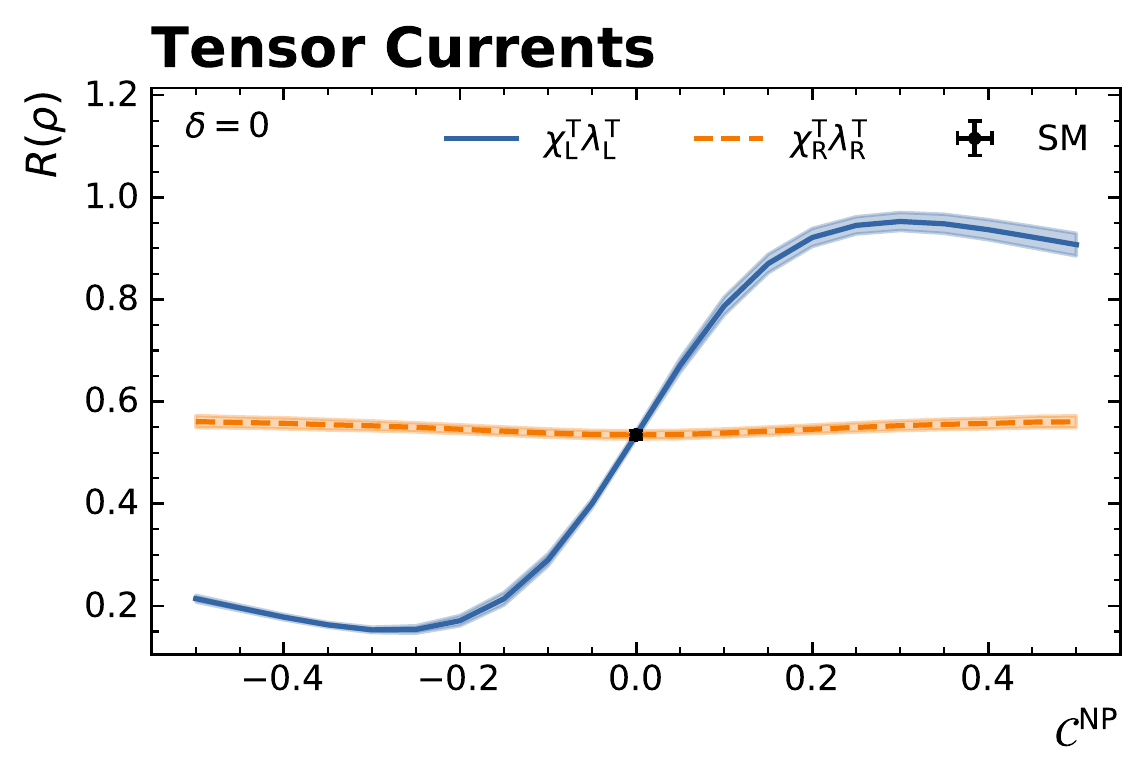}
\includegraphics[width=0.45\textwidth]{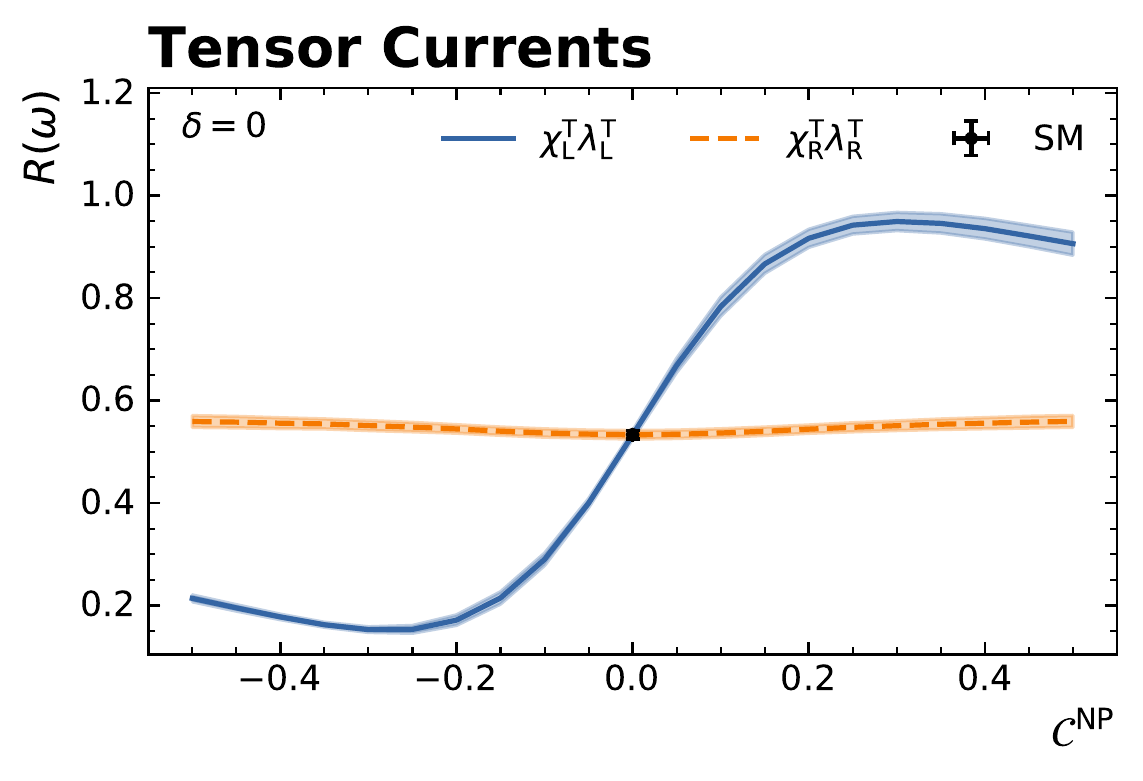}
\caption{The impact of a new physics current to the ratio of tauonic to leptonic final states in different new physics scenarios. The slight differences between the $\Prho$ and $\Pomega$ final states originates solely from the kinematic differences in the decay. The influence of new physics itself is the same, due to their identical vector-particle nature.}
\label{fig:new-physics-scans}
\end{figure}

The following predictions assume that new physics only contributes to the heavy leptons, i.e.\, the light leptons are free from any new physics contribution.
The complete basis of the four-Fermi operators mediating the $\Pbottom \rightarrow \Pquark \Ptau \Pnut$ is given by
\begin{equation}
2\,i\sqrt{2} V_\mathrm{ub} G_\mathrm{F}\, [\APquark \chi_j^i \gamma^\mu P_{j}\Pbottom] [\APlepton \lambda_l^k \gamma_\mu P_{l}\Pneutrino] \, ,
\end{equation}
where $\chi_j^i$ and $\lambda_l^k$ are the new physics coupling constants to the quark and lepton current, respectively. These new physics couplings are normalized to the SM coupling strengths.
Furthermore, the indices $j, l = \{L, R\}$ denote the helicity of the $\Pbottom$ quark and the neutrino. The indices $i, k =\{ S, V, T\}$ indicate that the current of the interaction is either a scalar, vector, or tensor current. 
The influence of the new physics contribution on $\RRho$ and $\ROmega$ is the same, except for mass differences, since both are vector-like particles. The influence of new physics on the ratio $R$ for each individual coupling is shown in Figure~\ref{fig:new-physics-scans}.

\section{Summary and Outlook}
We have demonstrated that we were able to improve the precision of BCL expansion coefficients by combining theory predictions with experimental measurements. 
This leads to more precise predictions of $\RRho$ and $\ROmega$ and $\redRRho$ and $\redROmega$, reducing the uncertainties by 20\% and 7\%, respectively, for $R$ and $\hat{R}$.  In addition, we investigated the impact of new physics contributions on $R$ for all four-Fermi operators. 

\bibliographystyle{JHEP}
\bibliography{skeleton}

\begin{thebibliography}{1}
\bibitem{NoComplexI}
{The helicity amplitudes are simplified by omitting the complex $i$}.
\end{thebibliography}

\end{document}